\documentclass[prl,letterpaper,aps,floatfix,twocolumn]{revtex4}
\usepackage{graphicx}
\usepackage{amsmath}
\usepackage{color}
\newcommand{\beq}{\begin{equation}}
\newcommand{\eeq}{\end{equation}}
\newcommand{\bk}{{{\bf{k}}}}

\newcommand{\br}{{{\bf{r}}}}

\newcommand{\bG}{{{\bf{G}}}}
\newcommand{\bA}{{\bf{A}}}

\newcommand{\ba}{{\bf{a}}}

\newcommand{\bb}{{\bf{b}}}

\newcommand{\bu}{{\bf{u}}}
\newcommand{\bm}{{\bf{m}}}

\newcommand{\beqa}{\begin{eqnarray}}
\newcommand{\eeqa}{\end{eqnarray}}

\begin{document}
\title{Continuous thermal melting of a two-dimensional Abrikosov vortex solid}
\author{J.~Iaconis, R.~G.~Melko, and A.~A.~Burkov}
\affiliation{Department of Physics and Astronomy, University of Waterloo, Waterloo, Ontario 
N2L 3G1, Canada}
\date{\today}
\begin{abstract}
We examine the question of thermal melting of the triangular Abrikosov vortex solid in two-dimensional superconductors 
or neutral superfluids.  We introduce a model, which combines lowest Landau level (LLL) projection with the magnetic Wannier basis 
to represent degenerate eigenstates in the LLL.  Solving the model numerically via large-scale Monte Carlo simulations, we find clear evidence for a continuous melting transition, in 
perfect agreement with the Kosterlitz-Thouless-Halperin-Nelson-Young theory and with recent experiments. 
\end{abstract}
\maketitle
The problem of melting of two-dimensional (2D) solids has been around for several decades. 
As was first pointed out by Kosterlitz and Thouless \cite{Thouless73}, a 2D solid to liquid transition 
can generally be expected to be very different from its three-dimensional (3D) counterpart. 
3D solids melt via a first-order transition, where the magnitude of the long-range crystal order, measured e.g.~by the strength of the Bragg peaks in the 
structure factor, drops to zero discontinuously from a finite value just below the melting temperature $T_m$.
In 2D,  long-range crystal order, which breaks continuous symmetry of spatial translations, is impossible at any nonzero 
temperature \cite{Mermin68}.  2D solids are thus characterized by power-law decay of crystalline correlations at low temperatures. 
This, however, is enough to give rise to a nonzero shear modulus and the low temperature phase is thus a true solid.
The order in this case is {\em topological}, in the sense that the solid phase is characterized by the absence of free topological defects, i.e.  
dislocations, which are bound into pairs with Burgers vectors equal in magnitude and opposite in direction. 
Kosterlitz and Thouless proposed \cite{Thouless73} that melting in 2D can happen continuously via the unbinding of dislocation pairs. 
Such a melting transition is then closely analogous to the well-known Kosterlitz-Thouless transition in 2D superfluids. 
Halperin, Nelson and Young \cite{HNY} developed the idea of Kosterlitz and Thouless into a detailed theory of dislocation-mediated 2D melting transition, which is now 
frequently referred to as KTHNY theory. 

Experimental confirmation of KTHNY theory has proven to be somewhat difficult to obtain \cite{Strandburg88}. 
However, by now there exist reports in the literature of apparently continuous melting transitions of triangular solids, which agree very well with KTHNY theory 
predictions \cite{Adams79}.    
In particular, very recently direct STM imaging of dislocation-mediated melting of a triangular vortex solid in a thin-film superconductor has 
reported a continuous transition \cite{Guillamon09}.
In contrast, there exists no direct theoretical evidence of a continuous finite-temperature melting transition in any microscopic model of 2D vortex solids \cite{Doniach79, firstorder, notransition}.

In this paper we address the problem of the melting of Abrikosov vortex lattices in 2D superconductors and neutral superfluids by introducing a model which combines lowest Landau level (LLL) projection with the magnetic Wannier basis to represent degenerate eigenstates in the LLL.
Solving the model using state of the art Monte Carlo simulations, we
obtain clear evidence for a continuous melting transition, in perfect agreement with KTHNY theory.

We start from the standard Ginzburg-Landau (GL) expression for the energy functional of a superconductor: 
\beqa
\label{eq:1}
{\cal H}&=&\int d^2 r \left[ \frac{1}{2 m^*} \left| \left(-i \hbar {\boldsymbol \nabla} + \frac{e^*}{c} \bA \right) \Psi(\br) \right|^2\right. \nonumber \\
&+& \left. a  |\Psi(\br)|^2 + \frac{b}{2} |\Psi(\br)|^4 \right].
\eeqa
The superconductor is placed in a perpendicular magnetic field ${\bf B} = {\boldsymbol \nabla} \times \bA$. 
We will assume that we are dealing with a strongly type-II superconductor, as is the case in the experiment of Ref.~\cite{Guillamon09}. 
In this case fluctuations of the electromagnetic field can be 
neglected and our results will then be applicable to neutral 2D superfluids in an artificial perpendicular magnetic field as well.
The quadratic part of Eq.~(\ref{eq:1}) is diagonalized by expanding the complex order parameter $\Psi(\br)$ in terms of Landau level eigenstates.
If the magnetic field is close to the upper critical field $H_{c2}$, or, equivalently, the superconducting coherence length is of the order 
of the magnetic length $\ell = \hbar c /e B$, the lowest Landau level (LLL) approximation can be used \cite{firstorder,notransition}, when only the contribution of the LLL to the eigenstate expansion of the order parameter is retained. We will adopt the LLL approximation henceforth. 
The order parameter is then written as:
\beq
\label{eq:2}
\Psi(\br) = \sum_\bm c_\bm \phi_\bm (\br).  
\eeq
Here $\bm$ is as yet unspecified LLL eigenstate label, $\phi_\bm(\br)$ is the corresponding eigenfunction and $c_\bm$ is the complex amplitude, 
corresponding to the LLL eigenstate $\phi_\bm(\br)$. 
A crucial ingredient of our work is a judicious choice of the LLL basis in Eq.(\ref{eq:2}), which should be chosen in such a way as 
to lead to the simplest representation of the low energy states of the system. 
As will be clear from the discussion below, the best LLL eigenstate basis for our problem is the basis of {\em magnetic Wannier functions}, 
first introduced in Ref.~\cite{Rashba97}, and employed e.g. in \cite{Burkov02,Burkov10}. 
We will only briefly mention the properties of magnetic Wannier functions that will be important for our present problem. 
Readers interested in a more detailed discussion should consult Refs.~\cite{Rashba97,Burkov02,Burkov10}. 
Magnetic Wannier states are defined on the sites of a triangular lattice with basis vectors $\ba_1= a \hat x,\,\, \ba_2 = a (\hat x + \sqrt{3} \hat y)/2$, so that 
$\br_\bm = m_1 \ba_1 + m_ 2 \ba_2$ with integer $m_{1,2}$. 
The lattice constant $a^2 = 4 \pi \ell^2 /\sqrt{3}$ is chosen in such a way that the unit cell of the lattice contains exactly one magnetic flux quantum 
$|\ba_1 \times \ba_2| = 2 \pi \ell^2$. The explicit form of magnetic Wannier functions $\phi_\bm(\br)$ will be unimportant for us, we will only mention that these states 
are normalizable but have a $1/r^2$ decay at long distances, which is the fastest decay compatible with the LLL projection. 
The utility of this set of LLL eigenstates for our problem becomes apparent when one notices that, unlike arbitrary spatial translations in the presence of a perpendicular magnetic field, translations by lattice vectors of the above lattice of  magnetic Wannier states commute with each other. This allows us to define lattice momentum $\bk$, belonging
to the first Brillouine zone of the triangular lattice and the corresponding Bloch states $\Psi_\bk(\br) = 1/ \sqrt{N_{\phi}} \sum_\bm \phi_\bm e^{\i \bk \cdot \br_\bm}$, 
where $N_{\phi}$ is the total number of magnetic flux quanta piercing the sample.  
As was shown in \cite{Burkov10}, the wavefunctions $\Psi_\bk(\br)$ are nothing but the magnetic Bloch states, first introduced by Eilenberger \cite{Eilenberger67}. 
These states form a complete orthonormal set of states in the LLL and correspond simply to triangular Abrikosov vortex lattices, with vortex cores located at:
\beq
\label{eq:3}
\br_{\bm \bk} = \br_\bm + (\ba_1+ \ba_2)/2 + \ell^2 \hat z \times \bk, 
\eeq
i.e. the Bloch momentum $\bk$ labels the center of mass positions of different Abrikosov lattice states. 
As shown in Ref.~\cite{Burkov10}, the GL functional Eq.~(\ref{eq:1}) takes a particularly simple form, when written in the magnetic Wannier basis. 
Namely, the quartic term of the GL functional $\int d^2 r |\Psi(\br)|^4 = \sum_{\bm_1 \ldots \bm_4} I_{\bm_1 \bm_2 \bm_3 \bm_4} c^*_{\bm_1} c^*_{\bm_2} c^{\vphantom *}_{\bm_3} c^{\vphantom *}_{\bm_4}$, where $I_{\bm_1 \bm_2 \bm_3 \bm_4} = \int d^2 r \phi^*_{\bm_1}(\br) \phi^*_{\bm_2}(\br) \phi^{\vphantom *}_{\bm_3}(\br) \phi^{\vphantom *}_{\bm_4}(\br)$,  
turns out to possess a low-energy symmetry corresponding to {\em center of mass conservation}, i.e. 
$I_{\bm_1 \bm_2 \bm_3 \bm_4} \sim \delta_{\bm_1 + \bm_2, \bm_3 + \bm_4}$. This symmetry is a reflection of the translational symmetry of the 2D plane, leading to 
the degeneracy of all the Abrikosov vortex lattice solutions, and of the LLL projection, which makes the set of Abrikosov vortex states a complete set. 
Taking into account that the matrix elements $I_{\bm_1 \bm_2 \bm_3 \bm_4}$ are short-range, namely have a $1/r^6$ decay at large distances, we then arrive 
at the following LLL representation of the GL functional:
\beq
\label{eq:4}
{\cal H} = -K \sum_P \cos(\theta_{1} - \theta_{2} + \theta_{3} - \theta_{4}), 
\eeq
where $P$ labels all possible smallest 4-site plaquettes of the triangular lattice (see Fig.~1) and we have taken $c_{\bm} \sim e^{i \theta_{\bm}}$, while 
$|c_{\bm}|^2$ is assumed fixed (note that this is not the same as neglecting fluctuations of the amplitude of $\Psi(\br)$, which would lead to Landau level
mixing). 
$K$ can be easily related to the parameters of the original GL functional Eq.(\ref{eq:1}), but will be left here as a phenomenological parameter. 
Its physical meaning, as will become clear shortly, is the {\em shear modulus} of the vortex lattice. 
Eq.(\ref{eq:4}) represents the shortest-range phase-dependent center of mass conserving quartic term on the triangular lattice. 
The quadratic term in the GL functional and the phase-independent quartic terms simply determine the magnitude of $|c_{\bm}|^2$ and consequently 
of the parameter $K$. 

Let us now demonstrate  that Eq.(\ref{eq:4}) indeed properly represents elasticity theory of the Abrikosov vortex solid. 
It is easy to show by direct substitution that the set of minimum energy states of Eq.(\ref{eq:4}) corresponds to all possible 
states with {\em uniform gradients} of the phase $\theta_{\bm}$ along the basis directions of the triangular lattice \cite{Balents03}.
Substituting the corresponding amplitudes $c_{\bm} \sim e^{i \theta_{\bm}}$ into Eq.(\ref{eq:2}) one obtains precisely the set of magnetic Bloch states $\Psi_{\bk}$, 
where each momentum $\bk$ corresponds to a given value of the phase gradient. 
It is then clear from Eq.(\ref{eq:3}) that in the long-wavelength elasticity theory of the vortex lattice, we can identify local vortex displacements with gradients of 
the phase $\theta$:
\beq
\label{eq:5}
\bu = \ell^2 \hat z \times {\boldsymbol \nabla} \theta.
\eeq
Defining the strain tensor $u_{ij} = (\partial_i u_j + \partial_j u_i)/2$ in the standard way and taking into account the incompressibility of the vortex lattice $u_{ii} = 0$ (summation over repeated indices is implicit), which immediately follows from (\ref{eq:5}), one obtains the following expression for the elastic energy:
\beq
\label{eq:6}
E = \mu \int d^2 r u_{ij} u_{ij} = \frac{\mu \ell^4}{2} \int d^2 r (\nabla^2 \theta)^2, 
\eeq
where $\mu$ is the shear modulus of the vortex lattice. 
To connect Eqs.(\ref{eq:4}) and (\ref{eq:6}), we expand the cosine in Eq.(\ref{eq:4}) to quadratic order in the phase, discard the ground state energy term,  and take the continuum limit:
\beq
\label{eq:7}
{\cal H} = \frac{9 K}{32} \frac{a^4}{ 2 \pi \ell^2} \int d^2 r (\nabla^2 \theta)^2 = \frac{3 \pi K \ell^2}{4} \int d^2 r (\nabla^2 \theta)^2, 
\eeq
which gives $K = 2 \mu \ell^2 / 3 \pi$ and defines the physical meaning of $K$. 
The lack of the $({\boldsymbol \nabla} \theta)^2$ term in (\ref{eq:7}) is a consequence of the center of mass conservation symmetry of 
Eq.(\ref{eq:6}). 
Eq.(\ref{eq:4}) can thus be thought of as a lattice regularization of the continuum elasticity theory of the incompressible Abrikosov vortex solid
Eq.(\ref{eq:6}). 
The representation of the continuum elasticity theory of the Abrikosov vortex lattice in terms of the phase Laplacian Eq.~(\ref{eq:6}) has in fact been 
obtained before by Moore \cite{Moore89}. 
It has not, however, been realized that a lattice regularization of (\ref{eq:7}), Eq.(\ref{eq:4}), can be used to study the melting transition. 

Before turning to numerical investigation of the phase diagram of the lattice model Eq.~(\ref{eq:4}), let us calculate some of the basic properties 
of the vortex solid phase using the continuum theory (\ref{eq:7}). 
In particular, correlations in the vortex solid phase can be quantified by two correlations functions: the phase correlation function
$G_{ph}(r) = \langle e^{i \theta(\br)} e^{-i \theta(0)} \rangle$ and the vortex position correlation function 
$G_{v}(r) = \langle e^{ i \bG \cdot [\bu(\br) - \bu(0)]} \rangle$, where $\bG$ is a reciprocal lattice vector. 
Straightforward calculation gives:
\beq
\label{eq:8}
G_{v}(r) \sim 1/r^{\eta_G},\,\, \eta_G = \frac{|\bG|^2 T}{4 \pi \mu},
\eeq
while $G_{ph}(r)$ vanishes in the thermodynamic limit for any nonzero $r$ and at any nonzero temperature. 
The exponent $\eta_G$ is in perfect agreement with the KTHNY results \cite{HNY}. 
The short-range nature of the phase correlation function $G_{ph}$ in the vortex solid has been emphasized by Te\v{s}anovi\'c \cite{Tesanovic94}, and our results 
are in agreement with Ref.~\cite{Tesanovic94}. 

As discussed above, we expect the power-law vortex solid order Eq.~(\ref{eq:8}) to exist up to a melting temperature $T_m$, at which free dislocations will 
appear and make the vortex positional correlations short range. 
To calculate $T_m$ we will employ the standard argument \cite{Thouless73}, comparing energetic and entropic contribution of a single dislocation to the 
free energy of the system. 
We assume the presence of an isolated dislocation with Burgers vector $\bb = a \hat x$:
\beq
\label{eq:9}
\oint {\boldsymbol \nabla} u_x \cdot d {\bf \ell} = a, 
\eeq
where the contour in the above integral encloses the dislocation core. 
Solution for the displacement field, minimizing the elastic energy of an incompressible solid Eq.~(\ref{eq:6}) under the constraint (\ref{eq:9}), is given by $u_x = a \phi/ 2 \pi + (a / 4 \pi) \sin(2 \phi), \,\, u_y = - (a / 4 \pi) \cos(2 \phi)$,
where $\phi$ is the polar angle \cite{Lubensky}.  
Substituting this solution into Eq.(\ref{eq:6}), one obtains the following result for the dislocation energy $E_d = (\mu a^2/ 2 \pi) \ln(L/a)$, where 
$L$ is the size of the system. 
Comparing this with the entropic contribution of the dislocation to the free energy $- T \ln(L/a)^2$, one obtains the melting
temperature \cite{HNY,Lubensky} $T_m = \mu a^2 / 4 \pi = \sqrt{3} \pi K / 2$, which gives:
\beq
\label{eq:12}
K/T_m = 2/ \sqrt{3} \pi.
\eeq
This can be compared with our numerical results, if $K$ is replaced by the actual value of the shear modulus at $T_m$ (see Eq.(\ref{eq:13}) below). 

According to KTHNY theory \cite{HNY}, the dislocation-mediated melting of a triangular solid results in an intermediate {\em hexatic phase}, between the solid and the isotropic liquid,  which has power-law orientational order of the solid, but no positional order.  
We do not expect to observe this phase in our numerics, which we will describe shortly. 
The reason is that our choice of the LLL basis and periodic boundary conditions that this choice implies, fixes the orientation of the vortex solid, thus precluding spontaneous orientational order.  This is true for the simulations of Ref.~\cite{firstorder}, which used the Landau gauge orbital basis, as well.  
We then expect to observe a single solid to liquid transition. The main question we would like to resolve is whether the transition is 
driven by the unbinding of dislocations and described by KTHNY theory, or it is first order as in the simulations of Ref.~\cite{firstorder} and thus possibly unrelated to unbinding of topological defects entirely. 

\begin{figure} {
\includegraphics[width=3 in]{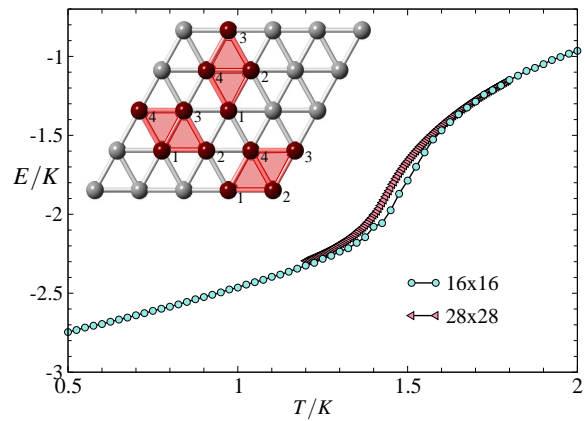} \caption{(Color online)
\label{Efig} 
The internal energy, $\langle {\cal H} \rangle$,  of Eq.~(\ref{eq:4}), using Monte Carlo simulations for two different system sizes.  The inset shows the labeling of sites on the three different plaquette orientations in  Eq.~(\ref{eq:4}).
}
} \end{figure}

We now turn to numerical Monte Carlo simulations of our lattice model Eq.~({\ref{eq:4}).  Taking the phases $\theta_{\bm}$ to be continuous variables between $0$ and $2 \pi$, we use a modified Metropolis algorithm that attempts an update on each phase variable selected randomly from a uniform distribution between $0$ and $\Delta \theta_{\rm max}$.  An important observation we make is that the choice of $\Delta \theta_{\rm max}$ critically affects the ergodicity of the update.  In conventional simulations of XY-type models,  $\Delta \theta_{\rm max}$ is varied as a function of temperature in order to maximize simulation acceptance rates.  We find that in our model this leads to problems with freezing into nearby metastable states.  In order to combat this, we systematically explored the simulation dynamics as a function of a temperature-independent $\Delta \theta_{\rm max}$.  We find that for insufficiently large $\Delta \theta_{\rm max}$ the simulation freezes into a metastable state of higher energy in an intermediate temperature regime,
however, with sufficiently large $\Delta \theta_{\rm max}$, the simulation is ergodic, and always finds the correct minimum free-energy state \footnote{We have confirmed this behavior using a more sophisticated and time-consuming parallel tempering simulation.}.
Figure~\ref{Efig} shows the simulation energy in the vicinity of a finite-temperature phase transition ($T_m \approx 1.3 K$, discussed below) for two different system sizes.  No evidence of a first-order discontinuity (latent heat) is apparent.

\begin{figure} {
\includegraphics[width=3.2 in]{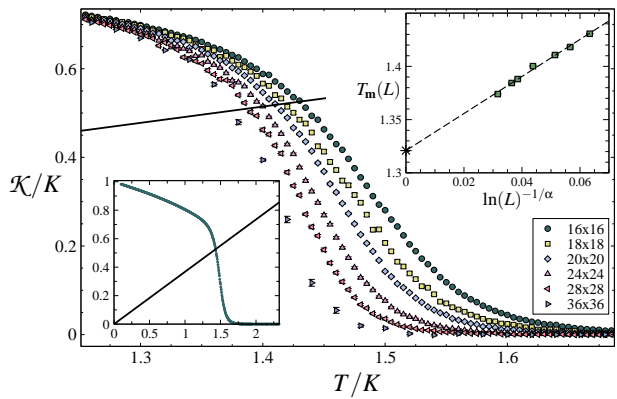} \caption{(Color online)
\label{hel}
The finite-size scaling behavior of the shear modulus through the finite-temperature transition.  The lower-left inset is data for linear lattice size $L=16$ over a larger temperature range.  The solid black line is the equation ${ \cal K} = (2/ \sqrt{3} \pi) K/T$, which from Eq.~(\ref{eq:12}) gives a finite-size estimate for $T_m(L)$.  The upper right inset is the finite-size scaling \cite{KTdoug} of $T_m(L)$, as discussed in text, which gives an estimate for $T_m(\infty)$ (the star) of $1.32 \, K$.
}
} \end{figure}

Figure \ref{hel} shows the main result of our simulations, which is the finite-size scaling behavior of the shear modulus ${\cal K}$, defined as the second derivative of the free energy with respect to the uniform shear angle 
$\varphi_s = \sqrt{3} \pi \ell^2 \nabla^2 \theta$:
\beq
\label{eq:13}
{\cal K} = \frac{1}{N_{\phi}} \left[\left \langle \frac{\partial^2 {\cal H}}{\partial \varphi_s^2} \right \rangle - \frac{1}{T} \left \langle 
\left(\frac{\partial {\cal H}}{\partial \varphi_s} \right)^2\right \rangle \right]_{\varphi_s = 0}, 
\eeq
where the prefactor in the expression for $\varphi_s$ is chosen so that ${\cal K}(T=0) = K$.
As is apparent from Fig.~\ref{hel}, we observe a single continuous phase transition from the vortex solid state with a nonzero shear modulus, to the high-temperature liquid phase.    Comparing the value of the ratio ${\cal K}(T_m)/T_m$ with the value in 
Eq.~(\ref{eq:12}), we see perfect agreement with the KTHNY theory.  We can make a highly-accurate quantitative assessment by comparing the finite-size scaling behavior of $T_m$ obtained using Eq.~(\ref{eq:12}) to the expected scaling form $T_m(L) - T_m(\infty) \propto  T_m(\infty) \ln(L)^{-1/\alpha}$ 
\cite{KTdoug} (Fig. \ref{hel} inset), where $\alpha = 0.36963$, from which we extract $T_m(\infty)/K  \approx 1.32$.
This scaling form follows from the KTHNY expression for the temperature dependence of the positional order correlation 
length $\xi \sim \exp[C/(T-T_m)^\alpha]$ \cite{HNY}. 

In conclusion, we have introduced a microscopic model for the melting of a 2D vortex solid using the magnetic Wannier basis 
to represent degenerate eigenstates in the LLL.  Solving the model using large-scale Monte Carlo simulations shows the finite-temperature melting transition is continuous,  in agreement with KTHNY theory and recent experiments.  We note that previous work on related model have observed a first-order melting transition \cite{firstorder}.  We believe that the most likely reason for this difference may be our neglect of the fluctuations of $|c_\bm|^2$, only retaining fluctuations of the 
phase $\theta_\bm$ (note again that this does not mean we are neglecting fluctuations of the amplitude of the order parameter $\Psi(\br)$).
The $|c_\bm|^2$ fluctuations, while not soft, could conceivably lead to a weak fluctuation-induced first order transition.
An indirect confirmation of this scenario is the fact that the melting temperature in Ref.~\cite{firstorder} is observed to be very close to the 
KTHNY melting temperature. 
Therefore, our result clearly demonstrates that dislocation unbinding is {\em the mechanism} behind the vortex lattice melting 
transition in 2D, even though the transition may be driven weakly first order in specific cases. 
}

\begin{acknowledgments}
We acknowledge useful discussions with A.H. MacDonald and A. Paramekanti. 
We thank the NSERC of Canada (JI, RGM, AAB) and the Ontario Ministry of Research and Innovation ERA (JI, RGM) for financial support.  
Computational facilities were provided by SHARCNET.
  
\end{acknowledgments}

\end{document}